\documentclass[11pt]{article}
\usepackage{amsmath,amsfonts}
\usepackage{amssymb}
\usepackage{latexsym}

\def \beq  {\begin{equation}}
\def \eeq  {\end{equation}}
\def \beqar {\begin{eqnarray}}
\def \eeqar {\end{eqnarray}}
\def\sqr#1#2{{\vcenter{\vbox{\hrule height.#2pt
\hbox{\vrule width.#2pt height#1pt \kern#1pt
\vrule width.#2pt}\hrule height.#2pt}}}}

\begin{document}
\def \CMP {{Commun. Math. Phys.}}
\def \PRL {{Phys. Rev. Lett.}}
\def \PL {{Phys. Lett.}}
\def \NPBProc {{Nucl. Phys. B (Proc. Suppl.)}}
\def \NP {{Nucl. Phys.}}
\def \RMP {{Rev. Mod. Phys.}}
\def \JGP {{J. Geom. Phys.}}
\def \CQG {{Class. Quant. Grav.}}
\def \MPL {{Mod. Phys. Lett.}}
\def \IJMP {{ Int. J. Mod. Phys.}}
\def \JHEP {{JHEP}}
\def \PR {{Phys. Rev.}}
\def \JMP {{J. Math. Phys.}}
\def \GRG{{Gen. Rel. Grav.}}
\fontfamily{cmr}\fontsize{11pt}{17.2pt}\selectfont
\begin{titlepage}
\null\vspace{-62pt} \pagestyle{empty}
\begin{center}
\vspace{1truein} 

 {\Large\bfseries
 {Infrared Modified Gravity with Dynamical Torsion}}

\vskip .1in
\vspace{.5in} {\large V. NIKIFOROVA$^{a,c}$, S. RANDJBAR-DAEMI$^b$,
V. RUBAKOV$^c$}~\footnote{E-mail:
{\fontfamily{cmtt}\fontsize{11pt}{15pt}\selectfont
seif@ictp.trieste.it, rubakov@ms2.inr.ac.ru}}\\
\vspace{.1in}{\itshape $^a$Physics Department, Moscow State University,
Moscow, Russia}\\
\vspace {.1in}{\itshape $^b$The Abdus Salam International Centre
for Theoretical Physics, Trieste, Italy}\\
\vspace{.1in}{\itshape $^c$Institute for Nuclear Research of the Russian Academy of
Sciences,
Moscow, Russia}\\

\vspace{.4in}
\centerline{\large\bf
}
\end{center}
We continue the recent  study of the possibility of
constructing a consistent infrared modification of gravity by
treating the vierbein and connection as independent
dynamical fields. We present  the generalized Fierz--Pauli
equation that governs the propagation of a massive spin-2
mode in a model of this sort in the backgrounds of arbitrary torsionless
Einstein manifolds.
We show explicitly that the number of propagating
degrees
of freedom in these backgrounds remains the same as in flat space-time.
This generalizes the recent result that
the Boulware--Deser phenomenon does not occur in de~Sitter and anti-de~Sitter
backgrounds.
We find that, at least for weakly curved
backgrounds,  there are  no ghosts in the model. We also briefly discuss
the interaction of sources in flat background.

\end{titlepage}
\pagestyle{plain} \setcounter{page}{2}

\section{Introduction and summary}
\label{sec:intro}
The possibility that gravity may get modified at large distances
attracts considerable interest, which is
 motivated, in particular, by the
accelerated expansion of the Universe. Deforming General Relativity
in the infrared is, however, not at all easy. The problems one has to
worry about are well understood in the context of the
Fierz--Pauli
theory~\cite{Fierz:1939ix} of massive graviton.
Already at the linearized level about
Minkowski
background, the Fierz--Pauli gravity
exhibits
the van~Dam--Veltman--Zakharov
discontinuity~\cite{vanDam:1970vg,Zakharov:1970cc},
which, taken at face value, implies that the theory deviates
from General Relativity even at short distances.
Below the Vainshtein radius, however, non-linear effects become
important~\cite{Vainshtein:1972sx} and, in fact, these effects may
cure the van Dam--Veltman--Zakharov problem~\cite{Babichev:2009us}.
The disaster occurs in curved backgrounds, where an extra,
Boulware--Deser  propagating mode
shows up~\cite{Boulware:1973my,Creminelli:2005qk,Deffayet:2005ys},
over and beyond the five modes of massive graviton. This extra mode has
ghost kinetic term and makes the theory unacceptable.


There are various theories that pretend to be free of at least some of
the above problems. These include theories which deform General Relativity
in backgrounds other than
%
Minkowski~\cite{Deser:2001pe,Deser:2001wx,Porrati:2001db}, theories
with extra
dimensions~\cite{Dvali:2000hr,deRham:2007xp,Kaloper:2007ap,Kaloper:2007qh,Kobayashi:2008jc}
(for a review see
Ref.~\cite{Gabadadze:2003ii}) and
theories with broken
Lorentz-invariance~\cite{Rubakov:2004eb,Dubovsky:2004sg,
Dubovsky:2004ud,Berezhiani:2007zf,Blas:2009my} (for a review see
Ref.~\cite{Rubakov:2008nh}).
Yet another option is to consider theories whose
independent fields are both vierbein and connection and whose Lagrangians
contain terms quadratic in curvature as well as mass terms for the
torsion field. Some theories of the latter type have both massless
and massive spin-2 modes in the spectrum about Minkowski background
and nevertheless are free of pathologies (ghosts or tachyons)
at the linearized level in this
background~\cite{Hayashi:1979wj,Hayashi:1980ir,Hayashi:1980qp,Sezgin:1979zf}.

It is worth noting that theories with dynamical vierbein and connection
may be viewed as gauge theories \footnote{Vierbein 
and connection as gauge fields of the Poincare group  
have been introduced by Kibble~\cite{Kibble:1961ba}
and studied extensively by Hehl and collaborators. 
For a historical review and references to earlier work see 
Ref.~\cite{Hehl:2007bn}.
Our point here is to study the possibility of a 
consistent infrared modification of gravity. }
with spontaneously
broken local Lorentz
invariance~\cite{Percacci:1984ai,Percacci:1990wy}. This viewpoint
opens up a possibility of unification of gravity with the Yang--Mills
theory~\cite{Dell:1986pw,Nesti:2007ka,Alexander:2007mt}. We will not
pursue this approach  and concentrate on degrees
of freedom describing dynamical space-time geometry itself.

Once
a geometrical theory has a massive spin-2 mode at the linearized level
in Minkowski background, a question arises of whether or not
this theory
has the Boulware--Deser mode in curved backgrounds.
Recently, this question has been addressed~\cite{Nair:2008yh}
in the context of one of the models of
Refs.~\cite{Hayashi:1979wj,Hayashi:1980ir,Hayashi:1980qp,Sezgin:1979zf}.
The analysis in Ref.~\cite{Nair:2008yh} has been restricted to
de~Sitter and anti-de~Sitter backgrounds. The result has been
encouraging, as it has been found that
there are no Boulware--Deser modes in these backgrounds.
In this paper we extend the analysis to arbitrary Einstein backgrounds
with vanishing torsion and show that the Boulware--Deser mode
does not appear in this more general case as well. If the curvature
of the background is sufficiently small, propagating modes are neither
ghosts nor tachyons. Hence, our analysis supports the conjecture
of Ref.~\cite{Nair:2008yh} that the model is healthy at least for
sufficiently weak fields.

As a by-product, we identify the field that reduces to the
massive spin-2 field as the curvature of the background is
switched off, and obtain the equation for this field that
serves as a   generalization of the Fierz--Pauli equation
to arbitrary Einstein backgrounds.

In the end of this paper we consider the interation of
sources in our model at the linearized level in Minkowski
background. We rederive the result
of
Refs.~\cite{Hayashi:1980ir}
that within the appropriate range of parameters, there are no ghosts
or tachyons. The interaction between symmetric and conserved
energy-momentum tensors is mediated by both massless and massive spin-2
fields, the relative strength depending on the parameters of
the model. In this sense our model is indeed an infrared-modified
gravity.
The interaction due to massive spin-2 field does exhibit
the van~Dam--Veltman--Zakharov discontinuity. We leave for future study
the questions of whether the Vainshtein phenomenon occurs in our
model, and if so, whether it cures the van~Dam--Veltman--Zakharov
problem.

The paper is organized as follows. We introduce the model,
its field equations and the Einstein manifold solutions in
section~\ref{sec:model}. In section~\ref{sec:vectors} we discuss the
behaviour of vector components of the torsion field in the
Einstein backgrounds. The results of  section~\ref{sec:vectors}
have been already obtained in Ref.~\cite{Nair:2008yh}; we present
them for completeness and later use. Section~\ref{sec:tensor}
contains the main results of this paper.
In section~\ref{subsec:genFP} we derive the equation that the massive 
tensor
field perturbations obey in the Einstein backgrounds. This equation
may be viewed as
a generalization of the Fierz--Pauli equation.
We then proceed to show that  in arbitrary Einstein backgrounds,
this equation describes five propagating
modes, the right number for massive spin-2 field. We begin in
section~\ref{subsec:counting} by counting the number of constraints
--- equations that contain time derivatives of at most first order ---
among the ten equations constituting the generalized
Fierz--Pauli system. We find that there are five such constraints,
which suggests that there are indeed five propagating modes.
In section~\ref{subsec:stuck}
we make use of the St\"uckelberg formalism to
confirm this result and show that at least
in weakly curved backgrounds,
neither of the propagating modes is a ghost.
We note, however, that the ghost problem may reappear in
backgrounds of sufficiently high curvature. Finally,
in section~\ref{sec:sources} we study the linearized theory
with sources in Minkowski background and obtain both the fields
induced by the sources and the action that describes the
interaction of the sources. Some of the results of this 
section are contained in Ref.~\cite{Hayashi:1980ir}; however,
our emphasis will be different. 
We conclude in section~\ref{sec:conclusions}
by discussing directions for future studies.

\section{The model}
\label{sec:model}

\subsection{Action}

 We follow the notations of
Refs.~\cite{Hayashi:1979wj,Hayashi:1980ir, Hayashi:1980qp,Nair:2008yh}
and denote the vierbein  by $e_\mu^i$ and the
 connection
by
 $A_{ij\mu}=-A_{ji\mu}$, where $\mu=0,1,2,3$ and  $i,j=0,1,2,3$
are the space-time and
tangent space indices, respectively.
We often use the tangent space basis, in which
the indices are
raised and lowered by the Minkowski metric $\eta_{ij}$.
The connection can be
viewed as an $O(1,3)$ gauge field. It is conveniently
decomposed as follows,
\beq
A_{ijk} \equiv A_{ij \mu} e^\mu_k = \frac{1}{2}\left(
T_{ijk} - T_{jik} - T_{kij} +
  C_{ijk} - C_{jik} - C_{kij} \right) \; ,
\label{T}
\eeq
where
\beq C_{ijk}=e_j^\mu e_k^\nu(\partial_\mu
e_{i\nu}-\partial_\nu e_{i\mu}) = - C_{ikj}
\nonumber
\eeq
is constructed from vierbein and $T_{ijk}=-T_{ikj}$ is the torsion tensor.
The latter can in turn be decomposed into its irreducible
 components under the  local $O(1,3)$ group,
\beq
 T_{ijk}=\frac{2}{3}(t_{ijk}-t_{ikj})+\frac{1}{3}(\eta_{ij}v_k
 -\eta_{ik}v_j)+\varepsilon_{ijkl}a^l
\label{T'}
\eeq
where the field $t_{ijk}$ is symmetric with respect to the
interchange of $i$ and $j$ and  satisfies the cyclic and trace
identities,
\beq
t_{ijk}+t_{jki}+t_{kij}=0, \quad\quad\quad
\eta^{ij}t_{ijk}=0,\quad\quad\quad \eta^{ik}t_{ijk}=0
\label{T''}
\nonumber
\eeq
The 24 independent components of $T_{ijk}$ break up into
4 components of $v_i$, 4 components of $a_i$ and
16 independent components of $t_{ijk}$.

The curvature, as usual in gauge theories, is defined by
\beq
 F_{ijmn}=e_{m}^{\mu}e_{n}^{\nu}(\partial_\mu A_{ij\nu}-\partial_\nu
 A_{ij\mu} +A_{ik\mu}{A^{k}}_{j\nu}-A_{ik\nu}{A^{k}}_{j\mu})
\nonumber
\eeq
The model studied in this paper, as well as in Ref.~\cite{Nair:2008yh},
is defined by the action
\beq
S = \int~d^4 x~e~ \left(L_F + L_T\right) \; ,
\nonumber
\eeq
where  $e= \det e_\mu^i$,
\beq
L_F= c_1F + c_2 +
c_3F_{ij}F^{ij}+c_4F_{ij}F^{ji}+c_5F^2+c_6(\varepsilon_{ijkl}F^{ijkl})^2
\nonumber
\eeq
and
\beq
L_T= \alpha\left(t_{ijk}t^{ijk} -v_iv^i+\frac{9}{4} a_ia^i\right) \; ,
\nonumber
\eeq
with
\beq
F_{jl}=
\eta^{ik}F_{ijkl},\quad\quad\quad F= \eta^{jk}F_{jk},\quad\quad\quad
\varepsilon \cdot F= \varepsilon_{ijkl}F^{ijkl} \; .
\label{contract}
\eeq
Here $c_1, \dots , c_6$ are ``coupling constants'' obeying, apart from
sign restrictions,
the only condition
%
\beq
c_3+c_4+3c_5=0 \; .
\nonumber
\eeq
In what follows, three combinations of these parameters will be used,
\begin{eqnarray}
\tilde{\alpha} &=& \alpha + \frac{2}{3} c_1
\\
\Lambda &=& - \frac{c_2}{6c_1} \; .
\label{Lambda-def} \\
\varkappa &=& 2\Lambda +\frac{\tilde{\alpha}}{2 c_5}
\label{kappa-def}
\end{eqnarray}
By appearance, the term $L_F$ has the form of  kinetic term for
the connection (plus the cosmological constant term $c_2$),
while $L_T$ is  torsion mass term.

We note in passing that the Lagrangian $L_T$ does not explicitly break
local $O(1,3)$ invariance, so the entire action is
invariant under both local frame rotations and general coordinate
transformations. For $c_2=0$, the model admits Minkowski space-time
as a solution of the field equations. In that case, the
local $O(1,3)$ invariance is spontaneously broken by the
background value of the vierbein field,
cf. Refs.~\cite{Percacci:1984ai,Percacci:1990wy,Dell:1986pw,Nesti:2007ka,Alexander:2007mt}.

The model with $c_2=0$
is free of ghosts and tachyons in Minkowski background
provided
the parameters
satisfy certain
inequalities~\cite{Hayashi:1979wj,Hayashi:1980ir,Hayashi:1980qp,Sezgin:1979zf},
which in our notations read
\beq
c_1>0 \;, \quad\quad\quad  c_5<0 \; , \quad\quad\quad
c_6>0 \; , \quad\quad\quad
\alpha<0 \; , \quad\quad\quad  \tilde\alpha>0 \; .
\label{cond}
\eeq
Non-vanishing value of $c_2$ enables one to have de~Sitter or
anti-de~Sitter solution with vanishing torsion in this model,
with cosmological constant equal to $\Lambda$. In the latter case,
the requirement of the absence of tachyons imposes one more
 condition~\cite{Nair:2008yh}, $c_5\varkappa > 0$, i.e.,
\beq
  \tilde{\alpha} > - 4\Lambda c_5 \; .
\label{may6-2}
\eeq
Since $c_5 < 0$, the latter condition is non-trivial for positive
$\Lambda$. Once the above conditions are satisfied, the theory
is healthy in de~Sitter and
anti-de~Sitter backgrounds~\cite{Nair:2008yh}.

\subsection{Field equations}

There are two sets of
field equations in our model. One consists of the gravitational field
equations obtained by varying the action with respect to vierbein,
%
\beqar &&c_1F_{ji}+ c_3(F^m{_i}F_{mj} -
{{F_{j}}^{mn}} _{i} F_{mn})+ c_4(F^m{_i}F_{jm} - {{F_{j}}^{mn}} _{i}
F_{nm})+ 2c_5F_{ji}F\nonumber \\
&&
+2c_6\varepsilon_{kmnj}{F^{kmn}}_i(\varepsilon_{rpqs}F^{rpqs})
+(D^k+v^k)F_{ijk}+ H_{ij}-\frac{1}{2}\eta_{ij}(L_F + L_T)=0
\label{Ein}
\eeqar
where \beqar
F_{ijk}&=&
\alpha\left[(t_{ijk}-t_{ikj})-(\eta_{ij}v_k
-\eta_{ik}v_j)-\frac{3}{4}\varepsilon_{ijkl}a^l\right]
\nonumber\\
H_{ij}&=&T_{mni}{F^{mn}}_j-\frac{1}{2}T_{jmn}{F_i}^{mn}
\nonumber
\eeqar
and $D_i$ is the covariant derivative with respect to the
connection $A_{ijk}$.
Note that these
equations have both symmetric and antisymmetric parts.
%

By varying the action with respect to
the connection $A_{ij\mu}$ one finds another set of equations,
\beqar
&&c_3\left\{\eta^{ik}(D_m+\frac{2}{3}v_m)F^{jm}
-\eta^{jk}(D_m+\frac{2}{3}v_m)F^{im}
-(D^i+\frac{2}{3}v^i)F^{jk}+(D^j+\frac{2}{3}v^j)F^{ik}\right\}\nonumber\\
&&+c_4\left\{\eta^{ik}(D_m+\frac{2}{3}v_m)F^{mj}
-\eta^{jk}(D_m+\frac{2}{3}v_m)F^{mi}
-(D^i+\frac{2}{3}v^i)F^{kj}+(D^j+\frac{2}{3}v^j)F^{ki}\right\}\nonumber\\
&&+c_5\left\{\eta^{ik}(D^j+\frac{2}{3}v^j)F
-\eta^{jk}(D^i+\frac{2}{3}v^i)F\right\}
+4c_6\left\{\varepsilon^{ijkm}(D_m+\frac{2}{3}v_m)
(\varepsilon \cdot F)\right\}\nonumber\\
&&-\left(\frac{4}{3}{t^k}_{[mn]}
+{\varepsilon^k}_{mnp}a^p\right)\left\{\phantom{\frac{}{}}
c_3(\eta^{im}F^{jn}-\eta^{jm}F^{in})+
c_4(\eta^{im}F^{nj}-\eta^{mj}F^{ni})\right. \nonumber\\
&&\left.+2c_5\eta^{im}\eta^{jn}F+
2c_6\varepsilon^{ijkm}(\varepsilon \cdot F)\right\}+H^{ijk}=0
\label{torsion}
\eeqar
where
\beq
H_{ijk}= -\tilde\alpha(t_{kij}-t_{kji})
+\tilde\alpha(\eta_{ki}v_j -\eta_{kj}v_i)
-\frac{3\tilde\alpha}{2}\varepsilon_{ijkl}a^l
\nonumber
\eeq
Equations (\ref{Ein}) and (\ref{torsion}) are not completely independent
because of the Bianchi identity.  In a theory with torsion,
this identity reads
\beq
D_k F_{ijlm} +
{T^n}_{kl}F_{ijmn} + {\rm cyclic}~ (klm)=0
\label{B1}
\eeq
Contracting this identity one obtains
\beq
D^{i} F_{ij} -\frac{1}{2}D_j F
= {T^i}_{kj} {F^{k}}_{i} +\frac{1}{2}{T^i}_{kl}{F^{kl}}_{ji}
\label{B2}
\eeq
We will see that these identities provide useful constraints in the
linearized theory.

\subsection{Einstein backgrounds}

Let us  consider torsion-free backgrounds.
For vanishing torsion, the curvature tensor and its contractions
reduce to  the Riemann tensor,
the Ricci tensor and the Ricci scalar,
$F_{ijkl}=R_{ijkl}, F_{ij}=R_{ij}$ and $F=R$,
respectively.
By inspecting eq.~(\ref{Ein}) one finds that it is satisfied for the
Einstein manifolds. The Riemann tensor for these manifolds has the following
form
\beq
R_{ijkl}=\Lambda(\eta_{ik}\eta_{jl}-\eta_{il}\eta_{jk}) +
W_{ijkl},
\nonumber
\eeq
where the Weyl tensor
$W_{ijkl}$ has all symmetries of the Riemann tensor
and is traceless in all pairs of
indices. One then has
\beq
R_{ij}=3\Lambda\eta_{ij} \; , \quad \quad\quad R=
12\Lambda \; .
\nonumber
\eeq
Using these properties one finds that eq.~(\ref{Ein})
reduces to the relationship  (\ref{Lambda-def})
between $\Lambda$ and coupling constants.
Equation (\ref{torsion}) is satisfied for the Einstein manifolds
automatically.

An important property of the Weyl tensor of the Einstein manifolds
follows from the Bianchi identity, namely
\beq
\nabla^i W_{ijkl}=0
\nonumber
\eeq
We will repeatedly make use of this property in what follows.

\section{Linearized theory in Einstein backgrounds: pseudovector
$a_i$ and vector $v_i$}
\label{sec:vectors}

One of the main purposes of this paper is to
study field perturbations about general
torsion-free Einstein backgrounds. The analysis of pseudovector $a_i$
and vector $v_i$ has been performed in Ref.~\cite{Nair:2008yh} where
is has been shown  that the vector field
$v_i$ does not have its own propagating modes, while the
pseudovector field
$a_i$ is a gradient, and its longitudinal part obeys the massive
Klein--Gordon equation. These properties
are exactly the same as
in the theory about Minkowski background.
For the sake of completeness and presentation of useful formulas,
we recapitulate the analysis
here.
%

We do not use special notation for the background
objects, unless there is a risk of an ambiguity;
the subscript $(1)$ refers to linearized perturbations.
The torsion components vanish for our backgrounds, and we do not
label their perturbations
by the subscript $(1)$.

To study the fields $v_i$ and $a_i$, it suffices to consider
certain combinations of the full equations (\ref{Ein})
and (\ref{torsion}). We begin with the antisymmetric part of the
gravitational equation (\ref{Ein}) whose complete form is
\beqar &&c_1 F_{[ji]}
+\frac{c_4}{2}(F^m{_i}F_{jm} - {{F^{m}}_{j}} F_{im})
-\frac{1}{2}({{F_j}^{mn}}_i-{{F_i}^{mn}}_j)(c_3F_{mn}+c_4F_{nm})\nonumber\\
&&\hskip .2in + 2c_5F_{[ji]}F+c_6(\varepsilon_{kmnj}{F^{kmn}}_i-
\varepsilon_{kmni}{F^{kmn}}_j)(\varepsilon_{rpqs}F^{rpqs}) \nonumber\\
&&\hskip .4in+(D^k+v^k)F_{[ij]k}+ H_{[ij]}=0 \; .
\label{AntEin}
\eeqar
By linearizing this equation about the Einstein background, one obtains
\beq
(c_1-4\Lambda c_3)F_{(1)[ji]}
-\nabla^kF_{(1)[ji]k}-(c_3-c_4){{W_j}^{[mn]}}_i F_{(1)[mn]}=0 \; .
\label{LinAnt}
\eeq
where  $\nabla$
denotes the covariant derivative with respect to the
background metric.
By linearizing $F_{ij}$ defined in (\ref{contract}) one finds
the following explicit expression for the antisymmetric part
\beq
 F_{(1)[ji]}=-\frac{2}{3\alpha}\nabla^kF_{[ji]k}
\label{may6-1}
\eeq
Hence,
eq.~(\ref{LinAnt}) is an algebraic equation for $F_{(1)[ij]}$
whose only solution is
\beq
F_{(1)[ij]}=0
\label{F1=0'}
\eeq
This result  simplifies all  other equations.

Now, let us write the curl of the torsion equation (\ref{torsion}).
Namely, we contract eq.~(\ref{torsion}) with $\varepsilon_{ijkl}$ to find
the complete curl equation,
\beqar
&&(c_3-c_4)\varepsilon_{lijk}D^iF^{jk}
-12c_6 D_l(\varepsilon \cdot F)-\frac{2}{3}\varepsilon_{ijkl}{t_n}^{ik}
(c_3F^{jn}+c_4F^{nj})-8c_6v_l (\varepsilon \cdot F)
\nonumber\\
&&-\frac{2}{3}(c_3-c_4)\varepsilon_{ijkl}v^iF^{jk}
-2(c_3F_{jl}+c_4F_{lj})a^j +\frac{2}{9}\tilde\alpha a_l=0
\label{curltorsion}
\eeqar
Substituting (\ref{F1=0'}) into eq.~(\ref{curltorsion}) and using the
fact that $\varepsilon \cdot F= 6\nabla^ia_i$ we obtain the following
linearized equation for $a_l$,
\beq 8c_6\nabla_l(\nabla \cdot a)-
\left( 2\Lambda c_5 +\frac{\tilde\alpha}{2} \right)a_l=0
\label{a}
\eeq
This equation shows that the pseudovector field  is a
gradient, $a_l = \nabla_l \sigma$.
Its longitudinal part obeys the Klein-Gordon equation
\beq \left(\nabla^2 -\frac{c_5 \varkappa}{8c_6}\right) \sigma=0
\label{aeq}
\eeq
where $\varkappa$ is defined in (\ref{kappa-def}).
The  spin zero field $\sigma$
has healthy kinetic term and  is not a tachyon
provided that the inequality (\ref{may6-2}) is satisfied.
The mass of this  field
coincides with the flat space result when $\Lambda=0$.

It is worth noting that using the explicit form of the right hand side of
eq.~(\ref{may6-1}) and
the fact that
$a_l$ is a gradient, one obtains from
eq.~(\ref{F1=0'})
%
the following constraint
\beq \nabla^k
t_{k[mn]}={\nabla_{[m}}v{_n}_]
\label{Const}
\eeq
This constraint already suggests that at least some components of
the field $v_i$ are not independent.

In fact, the entire field $v_i$ is not dynamical by itself, as it
can be expressed through  the tensor $t_{ijk}$.
To see this, one makes use of the trace of the torsion equation.
For obtaining its general form, one contracts eq.~(\ref{torsion})
with $\eta^{jk}$ and finds
\begin{align}
-3c_5&\left( D_jF^{(ij)}-\frac{1}{2}D^iF\right)
+(c_3-c_4)D_jF^{[ij]}-2c_5\left(v_jF^{(ij)}
-\frac{1}{2}v^iF\right)+(c_3-c_4)D_jF^{[ij]}
\nonumber\\
&+\frac{2}{3}(c_3-c_4)v_iF^{[ij]}
-3c_5t^{i(jn)}F_{(jn)}+\frac{1}{3}(c_3-c_4)t^{i[jn]}F_{[jn]}
\nonumber\\
&-\frac{1}{2}(c_3-c_4)\varepsilon^{ijnl}a_lF_{[jn]}
+6c_6a^i(\varepsilon \cdot F) + \frac{3}{2}\tilde\alpha v^i=0
\label{trtorsion}
\end{align}
Now, one
substitutes $F_{[ij]}=0$
in the linearized version
of this equation and finds
\beq
\left(D_jF^{ij}
-\frac{1}{2}D^i F\right)_{(1)}=
\varkappa v^i
\label{v}
\eeq
The left hand side of eq.~(\ref{v}) vanishes for vanishing torsion,
so it is proportional to torsion at the linearized level.
To obtain the explicit expression one makes
use of the Bianchi
identity (\ref{B2}).
Linearizing this identity
in the Einstein  background, one finds
\beq
\left( D^{i} F_{ij} -\frac{1}{2}D_j F \right)_{(1)}=
2\Lambda v_j +\frac{1}{2}{T^i}_{kl}{W^{kl}}_{ji}
\nonumber
\eeq
so that eq.~(\ref{v}) becomes
\beq
v_i= \frac{4c_5}{3\tilde\alpha}W_{ijkl}t^{j[kl]}
\label{v'}
\eeq
Hence, the vector field $v_i$ is not
an independently propagating field,
exactly as in the flat space.


\section{Spin-2 mode in Einstein backgrounds}
\label{sec:tensor}

\subsection{Generalized Fierz--Pauli equation}
\label{subsec:genFP}

Let us now derive the equations for propagating tensor
perturbations. To this end, we make use of the results of
section~\ref{sec:vectors} and write the liearized gravitational equation
(\ref{Ein}) in the following form,
\beq
c_1\left( F_{(1)ij}-\frac{1}{2}\eta_{ij}F_{(1)} \right) +\nabla^k F_{(1)ijk}+
3c_5W_{jmni}F_{(1)}^{mn}=0
 \label{symmet}
 \eeq
Hereafter we treat the field $F_{(1)ij}$ as well as the
combination $\nabla^kF_{(1)ijk}$ as symmetric with respect to
the interchange of indices $i$ and $j$, see eqs.~(\ref{may6-1}) and
(\ref{F1=0'}). The expression for $\nabla^k F_{(1)ijk}$ is, explicitly,
\beq
\nabla^kF_{(1)ijk}=
-\alpha\left[ 3\nabla^k t_{k(ij)}
-\frac{1}{2}  (\nabla_{i}v_j+\nabla_{j}v_i)+\eta_{ij}\nabla \cdot v
\right]
\label{S'}
\eeq
The remaining equation is the torsion equation (\ref{torsion}) whose
linearized version is
\begin{align}
\nabla_iF_{(1)jk} &-\nabla_jF_{(1)ik}
+\frac{1}{6}(\eta_{ik}\nabla_jF_{(1)}-\eta_{jk}\nabla_iF_{(1)})
\nonumber\\
&-\frac{1}{3}\left( 2\Lambda+
\frac{\tilde\alpha}{2c_5} \right)\left[
(\eta_{ik}v_j-\eta_{jk}v_i)+4t_{k[ij]}\right]=0
\label{spin2}
\end{align}
Here $v_i$ should be expressed in terms of $t_{ijk}$ according
to (\ref{v'}).

Equation~(\ref{spin2}) together with
eq.~(\ref{v}) may be used to express $t_{i[jk]}$
in terms of $F_{(1)ij}$,
\begin{align}
t_{k[ij]}=
\frac{1}{4\varkappa} & \left[
\left(3\nabla _i F_{(1)jk} - 3 \nabla_j F_{(1)ik} \right)
- \left(\eta_{ik} \nabla^l F_{(1)jl} -\eta_{jk} \nabla^l F_{(1)il}
\right) \right.
\nonumber\\
& \left.
+ \left(\eta_{ik } \nabla_j F_{(1)} - \eta_{jk}
\nabla_i F_{(1)} \right) \right] \; ,
\label{may7-1}
\end{align}
where $\varkappa$ is defined in (\ref{kappa-def}).
It is straightforward to check that eq.~(\ref{Const}),
with $t_{i[jk]} $ and $v_i$ expressed through $F_{(1)ij}$,
is identically satisfied.

Equation (\ref{may7-1}) determines, in fact, the full tensor $t_{ijk}$
in terms of $F_{(1)ij}$. Indeed, due to the identities
(\ref{T''}),
one has
\beq
t_{ijk}= \frac{2}{3}\left(t_{i[jk]}+t_{j[ik]}\right)
\label{dec19-1}
\eeq
Substituting this
into
eq.~(\ref{S'}) one obtains
\begin{align}
\nabla^l F_{jkl} =
- \frac{\alpha}{4\varkappa} &
\left[ 6 \nabla^2 F_{(1)jk} - 3 \left( \nabla_j \nabla^l F_{(1)kl}
+ \nabla^l \nabla_j F_{(1)kl}\right)  - 3 \left( \nabla_k \nabla^l F_{(1)jl}
+ \nabla^l \nabla_k F_{(1)jl}\right) \right.
\nonumber \\
& \left. + 6\eta_{jk}\nabla^m\nabla^n F_{(1)mn}
 - 4 \eta_{jk} \nabla^2 F_{(1)} +
4 \nabla_j \nabla_k F_{(1)} \right]
\label{may7-3}
\end{align}
Thus, equation~(\ref{symmet}) becomes the equation for the field
$F_{(1)ij}$,
\beq
\nabla^k F_{ijk} +
c_1\left(F_{(1)ij}-\frac{1}{2}\eta_{ij}F_{(1)}\right) +
3c_5W_{jmni}F_{(1)mn}=0 \; ,
\label{dec19-2}
 \eeq
where the first term is given by the right hand side of
(\ref{may7-3}).

Equation (\ref{dec19-2}) is a closed equation for the field
$F_{(1)ij}$, while the fields $t_{ijk}$ and $v_i$ are expressed
through $F_{(1)ij}$ according to
 eqs.~(\ref{may7-1}), (\ref{dec19-1}) and
(\ref{v'}). One can check that all equations of the linearized
theory are satisfied provided that the field $F_{(1)ij}$
obeys eq.~(\ref{dec19-2}).

Equation (\ref{dec19-2}) does not look similar to
the Fierz--Pauli equation
yet. To write it in a more familiar way, let us introduce
the field
\beq
 u_{ij} = F_{(1)ij} - \frac{1}{6} \eta_{ij} F_{(1)}
\nonumber
\eeq
Then eq.~(\ref{dec19-2}) becomes
\begin{align}
 \nabla^2 u_{ij} -\nabla^k\nabla_i u_{kj}
-&\nabla^k\nabla_j u_{ki}+\nabla_i\nabla_j u
+\eta_{ij}\left(\nabla^k\nabla^l u_{kl}-\nabla^2 u\right)
+6\Lambda\left(u_{ij}-\frac{1}{2}\eta_{ij}u\right)
\nonumber \\
&
- \left( 2\Lambda + \frac{2\varkappa c_1}{3\alpha}\right)(u_{ij}-\eta_{ij}u)
+\left(1-\frac{2\varkappa c_5}{\alpha}\right)W_{ilkj}u^{lk}=0
\label{FP``}
\end{align}
This equation reduces to the Fierz--Pauli equation
in  Minkowski background, with the mass of the spin-2 field given by
\beq
   m^2 = \frac{\tilde{\alpha} c_1}{3\alpha c_5} \; .
\label{may15-2}
\eeq
So,  eq.~(\ref{FP``}) may be viewed as the generalization of the
Fierz--Pauli equation to the Einstein backgrounds.

Equation~(\ref{FP``}) has particularly simple form in
de~Sitter and anti-de~Sitter
backgrounds
for which $W_{ilkj}=0$. In that case the trace and
divergence of this equation together with the equation
itself imply that $u_{ij}$ has to be traceless and divergence free.
We then obtain the Klein--Gordon equation
with the mass given by
\beq
M^2 = 4\Lambda\left(1+\frac{c_1}{3\alpha}\right)
+ \frac{\tilde\alpha c_1}{3\alpha c_5}
\nonumber
\eeq
in accord with Ref.~\cite{Nair:2008yh}. This mass obeys the
Higuchi bound~\cite{Higuchi:1989gz} provided the inequality
(\ref{may6-2}) is satisfied.

A remark is in order. Besides the tensor mode propagating
according to  eq.~(\ref{FP``}), there is of course a massless
tensor mode. The latter
corresponds to perturbations of the vierbein field
with vanishing torsion, and propagates according to the
linearized Einstein equations
 (written relative to an orthonormal basis of the background geometry),
$R_{(1)ij} = 0$. For this mode
$F_{(1)ij} = 0$, so  eq.~(\ref{FP``}) is trivially satisfied.

\subsection{ Counting the number of constraints}
\label{subsec:counting}

Let us show that out of
the ten equations satisfied by ten components of
$u_{ij}$,  five are constraints which involve at most
first order time derivatives. The remaining five
involve at most  second order time derivatives.
This suggests that the field $u_{ij}$ describes
five propagating modes, the right number for massive
spin-2 field.

Obtaining four out of five constraints is straightforward.
Indeed, the divergence of eq.~(\ref{FP``}) gives
first-order equations,
\beq
\tilde{M}^2\nabla^i( u_{ij}-\eta_{ij}u)
=\left( 1-\frac{2\varkappa c_5}{\alpha} \right)  W_{jikl}\nabla^k u^{il}
\label{tr'}
\eeq
where
\[
\tilde{M}^2 = M^2 - 2\Lambda
\]
These are obviously the four constraints.

To see that there are actually five constraints, let us
choose the coordinates in the  background manifold
such that the background metric components
are $g_{0a}=0= g^{0a}$, where $a=1,2,3$ and $g^{00}=g_{00}^{-1}$.
Working in coordinate basis rather than  orthonormal
one which we have used so far, we  perform the
$(3+1)$-decomposition of the field components $u_{\mu\nu}$
as well as the
 field equations. It is straightforward to see that  $(00)$-
as well as $(a0)$-components of eq.~(\ref{FP``})
involve at most first order
time derivatives. These are the four
constraints corresponding to eq.~(\ref{tr'}).
The
$(ab)$-components  of eq.~(\ref{FP``}) --- six equations --
are superficially second order in time.
However, one combination of the latter is in fact a constraint.
To see this, let us write the trace of eq.~(\ref{FP``}),
\beq
u=  \frac{2\varkappa c_5-\alpha}{\tilde{M}^2 \varkappa c_1}W_{imkn}
\nabla^i\nabla^n u^{mk}
\label{BD}
\eeq
To show that no time derivatives higher than
first order enter this equation, we write the right hand side of
this equation in an expanded form,
 \beq
 u= \frac{2\varkappa c_5-\alpha}{\tilde{M}^2 \varkappa c_1}
W_{\mu\lambda\sigma\nu}\nabla^\mu\nabla^\nu u^{\lambda\sigma}=
\frac{2\varkappa c_5-\alpha}{\tilde{M}^2 \varkappa c_1}
W^{0ab0}\nabla^2_0 u_{ab} +....
 \label{BD''}
 \eeq
where dots denote the terms which have
at most one time derivative. Now, we substitute
$\nabla^2_0 u_{ab}$ from eq.~(\ref{FP``}) into eq.~(\ref{BD''}).
One can show that the expression for
$\nabla^2_0 u_{ab}$ does not involve any term with more than
 one $\nabla_0$. Thus, the substitution shows
that eq.~(\ref{BD''}) is indeed another constraint reducing
the number of propagating modes from ten to five.

\subsection{St\"uckelberg treatment}
\label{subsec:stuck}

A simple way to
isolate the dangerous degrees of freedom is to make use of the
St\"uckelberg trick. As an example, this trick enables one to
see in rather straightforward manner how the van~Dam--Veltman--Zakharov
phenomenon and related effects emerge~\cite{Arkani-Hamed:2002sp}
 in the Fierz--Pauli theory in Minkowski background, and how
the Boulware--Deser
ghost mode appears~\cite{Creminelli:2005qk,Deffayet:2005ys}
in that theory in curved
backgrounds. Let us make use of the St\"uckelberg trick to see that
no Boulware--Deser mode is present in the theory with
the field equation (\ref{FP``}).

The quadratic action which corresponds to eq.~(\ref{FP``}) is,
up to an overall factor,
\beq
S= S_{inv}+S_m+S_W
\nonumber
\eeq
where
\begin{align}
S_{inv} &= \int d^4 x\sqrt{-g}\left\{
-\frac{1}{2} \nabla^k u_{ij}\nabla_k u^{ij} +  \nabla^k u_{ki}\nabla_l u^{li}
- \nabla^k u_{lk}\nabla^l u  +\frac{1}{2}\nabla_i u \nabla^i u \right.
\nonumber \\
& ~~~~~~~~~~~~~~~~~~~\left. -\Lambda \left(
u_{ij}u^{ij} +\frac{1}{2}u^2 \right) - W_{ilkj}u^{lk}u^{ij}\right\}
\label{Sinv}
\\
S_{m}&=  -\frac{\tilde{M}^2}{2}
\int d^4x\sqrt{-g}~
(u_{ij}u^{ij} -u^2)
\label{Sm}
\\
S_W&= s \int d^4x \sqrt{-g}~   W_{iklj}u^{kl}u^{ij}
\label{Sw}
\end{align}
Here
\beq
s= 1-\frac{2\varkappa c_5}{\alpha}
\nonumber
\eeq
The part (\ref{Sinv}) of the action
is invariant under the gauge transformation
\beq
u_{ij} = \bar{u}_{ij} + \nabla_i \zeta_j + \nabla_j \zeta_i
\label{may21-1}
\eeq
while the terms (\ref{Sm}) and (\ref{Sw}) are not invariant.
To implement the St\"uckelberg procedure, one introduces
new fields $\bar{u}_{ij}$,
$\xi_{i}$ and $\phi$ and writes
\beq
u_{ij} = \bar{u}_{ij} + \nabla_i \xi_j + \nabla_j \xi_i + \nabla_i
\nabla_j \phi
\nonumber
\eeq
Altogether, there are now 15 fields.
The theory is  invariant under two gauge symmetries,
\beq
\bar{u}_{ij} \to  \bar{u}_{ij} + \nabla_i \zeta_j + \nabla_j \zeta_i \; ,
\;\;\;\;\;\;
\xi_i \to \xi_i - \zeta_i
\nonumber
\eeq
and
\beq
\xi_i \to \xi_i + \nabla_i \psi  \; ,
\;\;\;\;\;\; \phi \to \phi - 2 \psi
\eeq
Provided that  all fields have
kinetic terms quadratic in derivatives,
these symmetries eliminate 10 degrees of freedom, and there remain
5 propagating modes.
It is not at all guaranteed, however, that the change of variables
(\ref{may21-1}) does not introduce higher-derivative
terms
in the action.
A counterexample is given by
the Fierz--Pauli gravity in curved backgrounds: the kinetic
term for the field $\phi$ has four derivatives (there are terms like
$(\Box \phi)^2$), and the sixths propagating mode, a ghost, appears in the
spectrum~\cite{Creminelli:2005qk,Deffayet:2005ys,Rubakov:2008nh};
this is the way the Boulware--Deser mode is seen in the
St\"uckelberg formalism.
Let us show that this phenomenon does not occur in our model.

The action $S_{inv} = S_{inv} (\bar{u})$
contains only the field $\bar{u}_{ij}$ and
is obviously second order in derivatives.
The change of variables (\ref{may21-1}) gives rise to the
derivative terms in the mass part of the action, 
\beq
S_m(\bar u, \xi, \phi)=-\frac{\tilde{M}^2}{2}\int d^4x \sqrt{-g}
\left\{
(\nabla_i\xi_j-\nabla_j \xi_i)(\nabla^i\xi^j
-\nabla^j \xi^i) - 2u \nabla^2 \phi - 3\Lambda \nabla_i \phi
\nabla^i \phi
+ \dots \right\}
\label{uxi}
\eeq
where the contribution proportional to $\Lambda$ appears in the process
of
integration by parts, and   dots denote terms with less than two
derivatives.
The term (\ref{Sw}) also contains derivatives, 
\beq
S_W (\bar u, \xi, \phi)= s\int  d^4 x \sqrt{-g}
W^{iklj} \left\{ 2\bar u_{ij}
\nabla_k \nabla_l \phi \right.
 \left. + 4 \left[\nabla_i(\xi_j + \nabla_j \phi)\right]
\left[\nabla_k (\xi_l + \nabla_l \phi)\right] + \dots\right\}
\label{barw}
\eeq
The part (\ref{uxi}) of the action gives healthy kinetic term for
the (gauge fixed)
vector field $\xi^i$ and provides kinetic mixing between
the fields $\phi$ and $u_{ij}$. This property is precisely the same
as in the Fierz--Pauli theory {\it in Minkowski
background}~\cite{Arkani-Hamed:2002sp}. The sum $(S_{inv} + S_m)$
can be diagonalized by the shift of the field
(cf. Ref.~\cite{Arkani-Hamed:2002sp}),
%

\beq
\bar{u}_{ij} \to \bar{u}_{ij} + \frac{\tilde{M}^2}{2} \eta_{ij} \phi
\nonumber
\eeq
As a result, the field $\phi$ obtains healthy kinetic term
\beq
S_{\phi} =  - \frac{3}{4} \tilde{M}^2 \left(\tilde{M}^2 - 2 \Lambda\right)
\int  d^4 x \sqrt{-g}~
\nabla_i \phi \nabla^i \phi
\nonumber
\eeq
The overall sign here is normal (non-tachyonic), provided that
inequality (\ref{may6-2}) is satisfied.

For the de~Sitter or anti-de~Sitter background one has $W_{ijkl}=0$,
so the St\"uckelberg action is clearly second order in derivatives
in that case and the fild $u_{ij}$ describes
five propagating modes,
in agreement with Ref.~\cite{Nair:2008yh}.

In the general Einstein background, the Weyl tensor does not vanish,
and the term (\ref{barw}) looks dangerous, as it appears to contain
four derivatives. However,  integrating by parts and using the properties
of the Weyl tensor, we write this term
in the following form, 
\beq
S_W (\bar u, \xi, \phi)= s\int  d^4 x \sqrt{-g}
W^{iklj} \left \{2\bar u_{ij}
\nabla_k \nabla_l \phi -  W_{iklm}(\nabla^m \phi\nabla^j\phi + 
4\nabla^m \phi\xi^j
 +2\xi_j\xi^j ) + \dots\right\}
\nonumber
\eeq
Hence, this term contains in fact at most two derivatives, so the
number of propagating modes remains equal to five. The Boulware--Deser
phenomenon is absent in our model, at least in the Einstein backgrounds.

In sufficiently weak background fields, when $|W_{ijkl}| \ll \tilde{M}^2$,
the mass term $S_{m}$ dominates over $S_W$, so the propagating modes are
not ghosts. In stronger background fields, the term $S_W$ induces explicit
two-derivative term in the action for the field $\phi$, and also
extra kinetic mixing between this field and $\bar{u}_{ij}$, so there
may appear ghost modes. We will comment on this point in
section~\ref{sec:conclusions}.

\section{Interaction between sources in Minkowski background}
\label{sec:sources}
Let us now consider the linearized theory in Minkowski background
and study the interaction between sources. Our purpose here is
twofold. First, we will confirm that all modes in this theory
linearized about flat space-time
are neither
ghosts nor tachyons. Second, we will see that the interaction between
conserved energy-momentum tensors which couple to the vierbein and do
not directly couple to connection is mediated by both massless
and massive spin-2 fields, so our model is indeed an infrared-modified
gravity.

Let us denote the sources coupled to the vierbein and connection by
$J_{i}^\mu$ and $S^{ij\mu}$, respectively, and introduce the source term
in the action,
\beq
S_{source}= \int d^4 x \left(
2 h^i_{\mu}J_i^{\mu} -\frac{1}{2} A_{ij\mu}S^{ij\mu}\right)
\label{f7}
\eeq
where $h^i_{\mu}$ is defined by
\beq
e^i_{\mu}= \delta^i_\mu + h^i_{\mu}
\nonumber
\eeq
We will still work with objects like $J^{ij}$, $S^{ijk}$,  defined
by $J^{ij} = J^{i\mu} \delta^j_\mu$, $S^{ijk} = S^{ij\mu} \delta^k_\mu$.
Note that the source $J^{ij}$ in general is not symmetric.

The theory is invariant
under linearized local frame rotations
and linearized general coordinate transformations. The
requirement that the source
term (\ref{f7}) respects these gauge symmetries gives two conservation
laws,
\beq
\partial^j J_{ij}= 0
\label{f4}
\eeq
and
\beq
\partial_l S^{ijl}=4J^{[ij]}
\label{f6}
\eeq
Note that eq.~(\ref{f4}) implies
\beq
\partial^j J_{(ij)}=-\partial^j J_{[ij]}
\label{may15-1}
\eeq
The local symmetries also enable one to set the antisymmetric part of
$h_{ij}$ equal to zero. Thus, in what follows we use the gauge
$h_{ij}=h_{ji}$.

Let us write down the linearized
field equations  with the sources.
We omit the subscript $(1)$ in this section.
The linearized gravitational equation is
  \beq
  c_1 \left( F_{ji}-\frac{1}{2}\eta_{ij}F \right) +\partial^k F_{ijk} =J_{ij}
  \label{f1}
  \eeq
Note that
the antisymmetric part of this equation gives
\beq
F_{[ij]}= -\frac{2}{3\alpha} J_{[ij]}
\label{f2}
\eeq
which is a constraint.
The propagation equation is thus the symmetric part of (\ref{f1}), namely,
\beq
c_1 \left( F_{(ji)} -\frac{1}{2}\eta_{ij}F \right)
+ \partial^k F_{(ij)k}=J_{(ij)}
\label{f3}
\eeq
The linearized torsion equations in the presence of source terms
have the following form,
\beqar
&& c_3\left (\eta^{ik} \partial_m F^{jm}  -
\eta^{jk} \partial_m F^{im}
- \partial^i F^{jk} + \partial^j F^{ik}\right)
\nonumber \\
&& +c_4\left( \eta^{ik} \partial_m F^{mj} - \eta^{jk} \partial_m F^{mi}
- \partial^i F^{kj} +\partial^j F^{ki} \right )
\nonumber \\
&&+
2c_5(\eta^ {ik} \partial^j F- \eta^ {jk} \partial^i F)
+ 4c_6 \varepsilon^{ijkm} \partial_m (\varepsilon \cdot F)
+ H^{ijk} =\frac{1}{2}S^{ijk}
\label{f5}
\eeqar
Making use of
(\ref{T}) and (\ref{T'})
one writes the source term (\ref{f7})
in terms of $h_{ij}$
and the components of the torsion.
The result is
\beq
S_{source}=
\int d^4 x \left( 2 h_{ij}\tau^{ij} +\frac{2}{3} t_{k[ij]}S^{ijk}
+\frac{1}{3}v_j {S^{ij}}_i -\frac{1}{4}\varepsilon_{ijkm} a^m S^{ijk}
\right)
\label{f9}
\eeq
where $\tau^{ij}$ is defined by
\beq
\tau^{ij}= J^{(ij)}- \frac{1}{2}\partial_m S^{m(ij)}
\label{current}
\eeq
In view of eqs.~(\ref{f6}) and (\ref{may15-1})
the source  $\tau_{ij}$ is conserved, as it should.
%

Let us also define the
following combinations of the sources,
\beq
S = \varepsilon_{ijkl} \partial_l S^{ijk}
\nonumber
\eeq
and
%
%
\beq
\sigma_{ij}= J_{(ij)} -\frac{\tilde\alpha}{\alpha}\frac{1}{2}\partial^m
S_{m(ij)}
\label{f24}
\eeq
It is now a matter of straightforward but tedious calculation to
find the solution to eqs.~(\ref{f1}) and (\ref{f5}).
The result is
\begin{align}
h_{ij}  = &\frac{1}{c_1}\frac{1}{k^2} \left(
\tau_{ij}-\frac{1}{2}\eta_{ij}\tau \right)
- \frac{\tilde\alpha}{c_1\alpha}\frac{1}{k^2+m^2}\left( \sigma_{ij}
-\frac{1}{3}\eta_{ij}\sigma \right)
\label{may21-2}
\\
v^i =& - \frac{1}{6\tilde\alpha} \left( {S^{ij}}_j
+ 8\frac{c_3}{c_1} ik_m \sigma^{mi} \right)
\\
a_l =& -\frac{1}{288 m_0 ^2 c_6} \frac{k_l S}{k^2+m_0 ^2}
+\frac{1}{18\tilde\alpha} \varepsilon_{ijkl}
\left\{ S^{ijk} +\frac { 2(c_3-c_4)}{3\tilde\alpha}k^ik_m S^{jkm} \right\}
\\
t_{k[ij]}&=\frac{i}{2\alpha}\frac{1}{k^2+m^2}\left\{k_i\left(\sigma_{jk}
-\frac{1}{3}\eta_{jk}\sigma\right)-k_j\left(\sigma_{ik}
-\frac{1}{3}\eta_{ik}\sigma\right)
+\frac{k_kk^m}{m^2}(k_i\sigma_{mj}-k_j\sigma_{mi})\right\}
\nonumber
\\
&
-
\frac{c_3-c_4}{36\tilde\alpha^2}k^m (k_iS_{jkm}-k_jS_{ikm}-2k_kS_{ijm})
-\frac{1}{12\tilde\alpha}(\eta_{ik}{S_{jm}}^m-\eta_{jk}{S_{im}}^m)
\nonumber
\\
&
-
\frac{1}{6\tilde\alpha}\{S_{ijk}+\frac{1}{2}(S_{ikj}-S_{jki})\}
+\frac{ic_3}{3c_1\tilde\alpha}k^m(\eta_{ik}\sigma_{mj}-\eta_{jk}\sigma_{mi})
\label{may21-5}
\end{align}
where $m_0$ is the flat limit of the
mass of the pseudoscalar
field (see eq.~(\ref{aeq})),
 \beq
 m_0^2= \frac{\tilde\alpha}{16 c_6}
\nonumber
 \eeq
and $m$ is given by (\ref{may15-2}).
Plugging these expressions back into the action
(this amounts to calculating $(1/2)$ of the source term (\ref{f9}))
we find the
action that describes the interaction of the sources.
%
We write its expression omitting the terms which are
ultra-local in the sources,
 \begin{align}
S_{int} = \int  d^4k &
 \left\{ \frac{1}{144\tilde\alpha}\frac{\bar S S}{k^2+m_0^2}
+\frac{1}{c_1} \; \frac{1}{k^2} \bar{\tau}_{ij}
\left(\tau^{ij}-\frac{1}{2}\eta^{ij}\tau\right) \right.
\nonumber\\
& \left.
-
\frac{\tilde\alpha}{\alpha c_1}\frac{1}{k^2+m^2}
\left[\bar\sigma_{ij} \left(\sigma^{ij}-\frac{1}{3}\eta^{ij}\sigma\right)
 + 2\frac{k^i k_m}{m^2}\bar\sigma_{ij} \left(\sigma^{jm}
-\frac{1}{3}\eta^{jm}\sigma\right)\right]\right\}
\label{summary}
\end{align}
where bar denotes complex conjugation. The three  terms here
correspond to exchange by massive spin-0 particle, massless spin-2
particle and massive spin-2 particle, respectively. The latter
exhibits the van~Dam--Veltman--Zakharov discontinuity, as is generally
the case. It is clear that with the restrictions on parameters
summarized in (\ref{cond}), neither of the modes is ghost or tachyon.

Of particular interest is the metric perturbation generated by the
symmetric energy-momentum tensor $\tau_{ij}= \tau_{(ij)}$
coupled to metric and not directly coupled
to torsion. In this case we have $J_{ij}=\sigma_{ij}=\tau_{ij}$, $S_{ijk}=0$,
$\partial_i \tau^{ij}=0$
and the expression (\ref{may21-2})
becomes
\beq
h_{ij}=
\frac{1}{c_1} \; \frac{1}{k^2} \left( \tau_{ij} -\frac{1}{2}\eta_{ij} \tau
\right) - \frac{\tilde\alpha}{\alpha c_1} \frac{
1}{k^2+m^2}\left( \tau_{ij}-\frac{1}{3}\eta_{ij} \tau \right)
\label{f19}
\eeq
Thus, in this case too, the interaction is mediated by both massless
and massive spin-2 fields, with relative strength being
a free parameter in our model. For completeness, let us write
down the linearized connection in this case,
\begin{align*}
A_{ijk} = & - \frac{i}{c_1} \frac{1}{k^2} \left\{
k_i \left( \tau_{jk} -\frac{1}{2}\eta_{jk} \tau
\right) - k_j \left( \tau_{ik} -\frac{1}{2}\eta_{ik} \tau
\right) \right\}
\\
& + \frac{i}{c_1} \frac{
1}{k^2+m^2}
 \left\{
k_i \left( \tau_{jk} -\frac{1}{3}\eta_{jk} \tau
\right) - k_j \left( \tau_{ik} -\frac{1}{3}\eta_{ik} \tau
\right) \right\}
\end{align*}
It is different from the Riemannian connection corresponding to
the vierbein perturbation (\ref{f19}) (see also the first line in
(\ref{may21-5})), which clearly shows mixing between vierbein and
torsion fields in our model.

\section{Conclusions}
\label{sec:conclusions}

The model discussed in this paper belongs to the class of
modified gravities in the sense that the gravitational force is
mediated by both massless and massive tensor fields.
Yet the model successfully passes a non-trivial consistency check:
in torsionless Einstein space backgrounds it has no pathologies in
the spectrum, at least for small enough background curvature.
Clearly, this model deserves further study.

One issue to be understood is whether the model is consistent in
more general backgrounds, including those with non-vanishing torsion.
Another is whether the Vainshtein mechanism cures the
van~Dam--Veltman--Zakharov problem; this issue can probably be
undersood in an appropriate decoupling limit, in analogy to
Refs.~\cite{Arkani-Hamed:2002sp,Nicolis:2004qq,Creminelli:2005qk,Deffayet:2005ys,Babichev:2009us}.
There is one more property of our model that may be related to
the Vainshtein non-linearity. We have seen in section~\ref{subsec:stuck} that
at the linearized level in the Einstein backgrounds, the longitudinal mode
$\phi$ may become a ghost for $|W_{ijkl}| > m^2$, where $m$ is the mass
of the spin-2 field. For spherical source, the Weyl tensor is
of order    $|W_{ijkl}| \sim R_S/r^3$, where $r$ is the distance
to the source and $R_S$ is the Schwarzschild radius.
Hence, the danger of a ghost occurs at
\beq
r \lesssim r_3 = \left( \frac{R_S}{m^2}\right)^{1/3}
\nonumber
\eeq
Note that $r_3$ is the smallest of all Vainshtein radii in the
Fierz--Pauli theories~\cite{Arkani-Hamed:2002sp}, the generic value being
\beq
r_V= r_5 =  \left( \frac{R_S}{m^4}\right)^{1/5}
\nonumber
\eeq
By analogy to other known infrared modified gravities we expect
that the longitudinal sector of our model goes non-linear at least
at $r \lesssim r_3$, so the Vainshtein mechanism may cure the
potential ghost problem as well.

The model we studied in this paper is geometrical, and its
full non-linear action is well defined. At the linearized level virtually
every source produces perturbations of both metric and torsion.
Thus, it
would be interesting to understand whether or not black
holes have similar property. If they do, the existence of
torsionless Ricci flat solution (for zero cosmological constant)
--- Schwarzchild black hole ---
would mean that black holes
in this model have torsion hair, and that the
Birkhoff's theorem is not valid. 

\section*{Acknowledgements}
We are indebted to V.P~Nair for stimulating discussions
and to
Stanley Deser for helpful correspondence.
S.R.D. is grateful to the participants of the first
$\Psi$G Workshop on the Consistent Modifications of Gravity,
especially to
Diego Blas, Cedric Deffayet, Gia Dvali and Arkady Vainshtein
for their criticism and useful comments. 
V.R. is indebted to R.P.~Woodard for useful discussion.
V.R.
has been supported in part by Russian Foundation for Basic
Research grant 08-02-00473.

\newpage


\begin{thebibliography}{99}

\bibitem{Fierz:1939ix}
  M.~Fierz and W.~Pauli,
  Proc.\ Roy.\ Soc.\ Lond.\  A {\bf 173}, 211 (1939).

\bibitem{vanDam:1970vg}
H.~van Dam and M.~J.~G. Veltman,
\newblock Nucl. Phys. {\bf B22}, 397 (1970).

\bibitem{Zakharov:1970cc}
V.~I. Zakharov,
\newblock JETP Lett. {\bf 12}, 312 (1970).


\bibitem{Vainshtein:1972sx}
A.~I. Vainshtein,
\newblock Phys. Lett. {\bf B39}, 393 (1972).

\bibitem{Babichev:2009us}
  E.~Babichev, C.~Deffayet and R.~Ziour,
  ``The Vainshtein mechanism in the Decoupling Limit of massive gravity,''
  arXiv:0901.0393 [hep-th]; \\
  E.~Babichev, C.~Deffayet and R.~Ziour, to appear.



\bibitem{Boulware:1973my}
  D.~G.~Boulware and S.~Deser,
  Phys.\ Rev.\  D {\bf 6}, 3368 (1972).

\bibitem{Creminelli:2005qk}
P.~Creminelli, A.~Nicolis, M.~Papucci and E.~Trincherini,
\newblock JHEP {\bf 09}, 003 (2005), [hep-th/0505147].



\bibitem{Deffayet:2005ys}
C.~Deffayet and J.-W. Rombouts,
\newblock Phys. Rev. {\bf D72}, 044003 (2005), [gr-qc/0505134].




\bibitem{Deser:2001pe}
S.~Deser and A.~Waldron,
\newblock Phys. Rev. Lett. {\bf 87}, 031601 (2001), [hep-th/0102166].

\bibitem{Deser:2001wx}
S.~Deser and A.~Waldron,
\newblock Phys. Lett. {\bf B508}, 347 (2001), [hep-th/0103255].

\bibitem{Porrati:2001db}
M.~Porrati,
\newblock JHEP {\bf 04}, 058 (2002), [hep-th/0112166].

\bibitem{Dvali:2000hr}
G.~R. Dvali, G.~Gabadadze and M.~Porrati,
\newblock Phys. Lett. {\bf B485}, 208 (2000), [hep-th/0005016].

\bibitem{deRham:2007xp}
  C.~de Rham, G.~Dvali, S.~Hofmann,
J.~Khoury, O.~Pujolas, M.~Redi and A.~J.~Tolley,
  Phys.\ Rev.\ Lett.\  {\bf 100}, 251603 (2008)
  [arXiv:0711.2072 [hep-th]].

\bibitem{Kaloper:2007ap}
  N.~Kaloper and D.~Kiley,
  JHEP {\bf 0705}, 045 (2007)
  [arXiv:hep-th/0703190].


\bibitem{Kaloper:2007qh}
  N.~Kaloper,
  Mod.\ Phys.\ Lett.\  A {\bf 23}, 781 (2008)
  [arXiv:0711.3210 [hep-th]].

\bibitem{Kobayashi:2008jc}
  T.~Kobayashi,
  Phys.\ Rev.\  D {\bf 78}, 084018 (2008)
  [arXiv:0806.0924 [hep-th]].

\bibitem{Gabadadze:2003ii}
G.~Gabadadze, ``ICTP lectures on large extra dimensions,''
[hep-ph/0308112].

\bibitem{Rubakov:2004eb}
  V.~A.~Rubakov,
  ``Lorentz-violating graviton masses: Getting around ghosts, low strong
  coupling scale and VDVZ discontinuity,''
  arXiv:hep-th/0407104.



\bibitem{Dubovsky:2004sg}
  S.~L.~Dubovsky,
  JHEP {\bf 0410}, 076 (2004)
  [arXiv:hep-th/0409124].

\bibitem{Dubovsky:2004ud}
  S.~L.~Dubovsky, P.~G.~Tinyakov and I.~I.~Tkachev,
  Phys.\ Rev.\ Lett.\  {\bf 94}, 181102 (2005)
  [arXiv:hep-th/0411158].

\bibitem{Berezhiani:2007zf}
  Z.~Berezhiani, D.~Comelli, F.~Nesti and L.~Pilo,
  Phys.\ Rev.\ Lett.\  {\bf 99}, 131101 (2007)
  [arXiv:hep-th/0703264].

\bibitem{Blas:2009my}
  D.~Blas, D.~Comelli, F.~Nesti and L.~Pilo,
  ``Lorentz Breaking Massive Gravity in Curved Space,''
  arXiv:0905.1699 [hep-th].



\bibitem{Rubakov:2008nh}
V.~A.~Rubakov and P.~G.~Tinyakov,
  Phys.\ Usp.\  {\bf 51}, 759 (2008)
  [arXiv:0802.4379 [hep-th]].

\bibitem{Hayashi:1979wj}
  K.~Hayashi and T.~Shirafuji,
  Prog.\ Theor.\ Phys.\  {\bf 64}, 866 (1980)
  [Erratum-ibid.\  {\bf 65}, 2079 (1981)].

\bibitem{Hayashi:1980ir}
  K.~Hayashi and T.~Shirafuji,
  Prog.\ Theor.\ Phys.\  {\bf 64}, 1435 (1980)
  [Erratum-ibid.\  {\bf 66}, 741 (1981)].

\bibitem{Hayashi:1980qp}
  K.~Hayashi and T.~Shirafuji,
  Prog.\ Theor.\ Phys.\  {\bf 64}, 2222 (1980).




\bibitem{Sezgin:1979zf}
  E.~Sezgin and P.~van Nieuwenhuizen,
  Phys.\ Rev.\  D {\bf 21}, 3269 (1980).

\bibitem{Kibble:1961ba}
  T.~W.~B.~Kibble,
  J.\ Math.\ Phys.\  {\bf 2}, 212 (1961).

\bibitem{Hehl:2007bn}
  F.~W.~Hehl and Y.~N.~Obukhov,
  ``Elie Cartan's torsion in geometry and in field theory, an essay,''
  arXiv:0711.1535 [gr-qc].


\bibitem{Percacci:1984ai}
  R.~Percacci,
  Phys.\ Lett.\  B {\bf 144}, 37 (1984).

\bibitem{Percacci:1990wy}
  R.~Percacci,
  Nucl.\ Phys.\  B {\bf 353}, 271 (1991)
  [arXiv:0712.3545 [hep-th]].

\bibitem{Dell:1986pw}
  J.~Dell, J.~L.~deLyra and L.~Smolin,
  Phys.\ Rev.\  D {\bf 34}, 3012 (1986).

\bibitem{Nesti:2007ka}
  F.~Nesti and R.~Percacci,
  J.\ Phys.\ A  {\bf 41}, 075405 (2008)
  [arXiv:0706.3307 [hep-th]].

\bibitem{Alexander:2007mt}
  S.~H.~S.~Alexander,
  ``Isogravity: Toward an Electroweak and Gravitational Unification,''
  [arXiv:0706.4481 [hep-th]].

\bibitem{Nair:2008yh}
  V.~P.~Nair, S.~Randjbar-Daemi and V.~Rubakov,
  ``Massive Spin-2 fields of Geometric Origin in Curved Spacetimes,''
  [arXiv:0811.3781 [hep-th]].


\bibitem{Higuchi:1989gz}
  A.~Higuchi,
  Nucl.\ Phys.\  B {\bf 325}, 745 (1989).

\bibitem{Arkani-Hamed:2002sp}
N.~Arkani-Hamed, H.~Georgi and M.~D. Schwartz,
\newblock Ann. Phys. {\bf 305}, 96 (2003), [hep-th/0210184].

\bibitem{Nicolis:2004qq}
  A.~Nicolis and R.~Rattazzi,
  JHEP {\bf 0406}, 059 (2004)
  [arXiv:hep-th/0404159].




\end{thebibliography}
\end{document}